\documentclass[
]{ceurart}

\usepackage{listings}
\usepackage{subcaption}
\usepackage[table,xcdraw]{xcolor}
\usepackage{makecell}
\usepackage{threeparttable}
\begin{document}

%%
%% Rights management information.
%% CC-BY is default license.
\copyrightyear{2026}
\copyrightclause{Copyright for this paper by its authors.
  Use permitted under Creative Commons License Attribution 4.0
  International (CC BY 4.0).}

%%
%% This command is for the conference information
\conference{Proceedings of the 1st Late Interaction Workshop (LIR) @ ECIR 2026, April 02, 2026.}

%%
%% The "title" command
\title{Nemotron ColEmbed V2: Top-Performing Late Interaction Embedding Models for Visual Document Retrieval}

%%
%% The "author" command and its associated commands are used to define
%% the authors and their affiliations.
\author[1]{Gabriel de Souza P. Moreira}[%
%orcid=0000-0002-9745-5014,
email=gmoreira@nvidia.com,
]
%\cormark[1]
\fnmark[1]
\address[1]{NVIDIA}

\author[1]{Ronay Ak}[%
%orcid=xxx,
email=ronaya@nvidia.com,
]
\fnmark[1]
%\address[3]{NVIDIA, Florida, USA}

\author[1]{Mengyao Xu}[%
%orcid=xxx,
%email=mengyaox@nvidia.com,
]
%\address[2]{NVIDIA, Santa Clara, USA}

\author[1]{Oliver Holworthy}[%
%orcid=xxx,
%email=oholworthy@nvidia.com,
]
%\address[4]{NVIDIA, London, UK}

\author{Benedikt Schifferer}[%
%orcid=xxx,
%email=bschifferer@nvidia.com,
]
\cormark[1]
%\address[4]{X, X, X}

\author[1]{Zhiding Yu}[%
%orcid=xxx,
%email=X,
]

\author[1]{Yauhen Babakhin}[%
%orcid=xxx,
%email=ybabakhin@nvidia.com,
]
%\address[4]{NVIDIA, XXX, XX}

\author[1]{Radek Osmulski}[%
%orcid=xxx,
%email=rosmulski@nvidia.com,
]
%\address[4]{NVIDIA, XXX, XX}

\author[1]{Jiarui Cai}[%
%orcid=xxx,
%email=jiaruic@nvidia.com,
]
%\address[4]{NVIDIA, XXX, XX}

\author[1]{Ryan Chesler}[%
%orcid=xxx,
%email=rchesler@nvidia.com,
]
%\address[4]{NVIDIA, XXX, XX}

\author[1]{Bo Liu}[%
%orcid=xxx,
%email=boli@nvidia.com,
]
%\address[4]{NVIDIA, New Jersey, USA}

\author[1]{Even Oldridge}[%
%orcid=xxx,
%email=eoldridge@nvidia.com,
]
%\address[4]{NVIDIA, Vancouver, Ca}

%% Footnotes
\fntext[1]{These authors contributed equally.}
\cortext[1]{Work done while working at Nvidia.}

%%
%% The abstract is a short summary of the work to be presented in the
%% article.
\begin{abstract}
Retrieval-Augmented Generation (RAG) systems have been popular for generative applications, powering language models by injecting external knowledge. Companies have been trying to leverage their large catalog of documents (e.g. PDFs, presentation slides) in such RAG pipelines, whose first step is the retrieval component. Dense retrieval has been a popular approach, where embedding models are used to generate a dense representation of the user query that is closer to relevant content embeddings. More recently, VLM-based embedding models have become popular for visual document retrieval, as they preserve visual information and simplify the indexing pipeline compared to OCR text extraction.

Motivated by the growing demand for visual document retrieval, we introduce Nemotron ColEmbed V2, a family of open-weights models that achieve state-of-the-art performance on the ViDoRe benchmarks. We release three model sizes—with 3B, 4B, and 8B parameters—based on pre-trained VLMs: NVIDIA Eagle 2 with Llama 3.2 3B backbone, Qwen3-VL-4B-Instruct and Qwen3-VL-8B-Instruct, respectively. The 8B model ranks first on the ViDoRe V3 leaderboard as of February 03, 2026, achieving an average NDCG@10 of 63.42.

We describe the main techniques used across data processing, training, and post-training—such as cluster-based sampling, hard-negative mining, bidirectional attention, late interaction, and model merging—that helped us build our top-performing models. We also discuss compute and storage engineering challenges posed by the late interaction mechanism and present experiments on how to balance accuracy and storage with lower dimensional embeddings.

\end{abstract}

%%
%% Keywords. The author(s) should pick words that accurately describe
%% the work being presented. Separate the keywords with commas.

%%
%% This command processes the author and affiliation and title
%% information and builds the first part of the formatted document.
\maketitle

\section{Introduction}

Retrieval-Augmented Generation (RAG) has become a widely adopted paradigm for enhancing language models generation with external knowledge, enabling them to retrieve and reason on relevant content from large-scale corpora. Numerous high-performing text retrieval models including NV-Embed~\cite{nv-embed},  NV-Retriever~\cite{nv-retriever}, Qwen3-Embedding~\cite{qwen3-embed}, and e5-mistral~\cite{e5-mistral} have been proposed, and evaluated on benchmarks such as MTEB~\cite{mteb,mtebnew} text benchmarks, that assume clean and well-formatted textual inputs. In contrast, real-world use cases typically involve documents stored in formats like PDFs, PowerPoint slides, or Word documents, requiring preprocessing pipelines to extract textual content (e.g. text parsing, OCR). This process often results in the loss of critical visual information for modalities like tables, charts, and infographics. To address those limitations, Visual Document Retrieval (VDR)~\cite{colpali} has been proposed to retrieve document pages directly from their image, with no need for text extraction, further preserving visual information and simplifying the indexing and search pipelines from document retrieval systems. If page text is easily available, both page image and text can be used for better page multimodal representation and retrieval.

Recent Vision-Language models (VLMs) aim to bridge the gap between text and image understanding by learning joint representations across modalities. Models such as Qwen-VL~\cite{qwenvl}, LLaMA-3.1-Nemotron-Nano-VL~\cite{nemotron}, PaliGemma 2~\cite{paligemma2}, NVIDIA’s Eagle 2~\cite{eagle, eagle25} and Nemotron Nano V2 VL~\cite{deshmukh2025nvidia} have demonstrated strong performance across a range of vision-language tasks. VLMs use image encoders like CLIP~\cite{clip}, SigLIP~\cite{siglip} and C-RADIO~\cite{cradio} to extract image features and project them to the LLM space. 
VLM decoder models have been adapted as contrastive embedding models for visual document retrieval, like Jina CLIP~\cite{xiao2024jina} and Nomic Embed Vision~\cite{nussbaum2024nomic}. 

ColBERT~\cite{colbert} demonstrated that a multi-vector late interaction approach could boost retrieval accuracy, for allowing deeper interaction between query and textual context tokens, compared to the pooling approaches (e.g., average, last) that compress the representation to a single-vector embedding. ColPali~\cite{colpali} VLM embedding model leveraged late interaction between text-image tokens for improved retrieval accuracy.

In order to evaluate visual document retrieval models, several benchmarks were  introduced. The most popular ones belong to the ViDoRe, which was released in three versions: V1~\cite{colpali},  V2~\cite{vidore2} and V3~\cite{loison2026vidorev3}. The latest Vidore V3~\cite{loison2026vidorev3} is an important expansion for evaluating VDR on complex real-world scenarios, including multi-type and multi-language queries across ten professional domains.

In this paper, we introduce the Nemotron ColEmbed V2, a family of state-of-the-art embedding models for visual document retrieval. Our best-performing model \emph{nemotron-colembed-vl-8b-v2} achieves an NDCG@10 of 63.42 on the Vidore V3 benchmark (+3\% to the second place), ranking first on that benchmark as of Feb. 03, 2026. 

We initialized our 3B model from NVIDIA's Eagle 2 vision-language model~\cite{eagle, eagle25} with Llama 3.2 3B LLM backbone and initialize our 4B and 8B models from Qwen3-VL
models~\cite{bai2025qwen3vltechnicalreport}. We also replaced the original causal attention with bidirectional attention, and fine-tuned the models through contrastive training with late interaction mechanism on curated datasets for visual document retrieval. Our training datasets contain both text-only and text-image examples for contrastive learning, and we apply hard negative mining following the methods proposed in NV-Retriever~\cite{nv-retriever} to improve retrieval accuracy. Finally, we use model merging to provide ensemble-level accuracy in a single model.

The main contributions of this paper are: 

\begin{itemize}
    \item We release three state-of-the-art models for visual document retrieval: \emph{llama-nemotron-colembed-vl-3b-v2}\footnote{\url{https://huggingface.co/nvidia/llama-nemoretriever-colembed-3b-v2}}, \emph{nemotron-colembed-vl-4b-v2}\footnote{\url{https://huggingface.co/nvidia/nemotron-colembed-vl-4b-v2}}, and \emph{nemotron-colembed-vl-8b-v2}\footnote{\url{https://huggingface.co/nvidia/nemotron-colembed-vl-8b-v2}}. The 8B model achieves top-1 performance in the MTEB ViDoRe V3 leaderboard, while the 4B and 3B models are among top-6, outperforming the models of the same size in the leaderboard on NDCG@10.
    \item We describe the techniques that helped boost our model accuracy to the top of ViDoRe leaderboards regarding data preprocessing (clustering-based sampling, hard-negative mining, cross-lingual translation), two-stage training, late interaction, and model merging.
    %\item We explore two-stage training strategy, where the first stage leverages large-scale text-only data while the second stage uses text-image data. Our results demonstrate that pretraining on large-scale text-only retrieval data significantly enhances the model’s performance on downstream text-image retrieval tasks, highlighting the transferability of textual retrieval capabilities to multimodal settings.
    \item Finally, we discuss some performance trade-offs of using late interaction mechanism, as its higher accuracy comes at the price of increasing indexing embeddings storage and higher compute at serving compared to simpler vector pooling. We provide an ablation on the reduced embedding sizes for the late interaction.
\end{itemize}
\section{Background}
\subsection{Visual Document Retrieval}

Visual document retrieval has transitioned from rigid pipelines to integrated multimodal frameworks. Early systems \cite{xu2020layoutlm, huang2022layoutlmv3} relied on OCR-centric pipelines, where text was first extracted and then processed by layout-aware encoders that fused content with 2D positions. Although effective for document understanding, these early designs were computationally prohibitive for first-stage retrieval and were largely restricted to reranking small candidate sets. In parallel, early CLIP-style bi-encoders \cite{ma2024unifying} offered global multimodal representations, but frequently struggled with cluttered pages where relevant information was sparse or layout-dependent.

The rapid advancement of Large VLMs has blurred the boundaries between OCR, layout detection, and visual understanding. Modern VLMs, trained on web-scale data, implicitly recognize rendered text and layout structures directly from pixels, eliminating the need for explicit OCR or handcrafted features. This shift has catalyzed a new generation of visual document retrieval models that combine VLM backbones with ColBERT-style late interaction \cite{colbert}. Notable implementations include ColPali \cite{colpali} (PaliGemma-based \cite{beyer2024paligemma}), ColQwen \cite{colpali} (Qwen2-based \cite{team2024qwen2}), Jina Embedding Models \cite{gunther2025jina} (Qwen2.5-based \cite{bai2025qwen25}), and NVIDIA’s Nemoretriever Colembed \cite{xu2025llama} (Eagle2-based \cite{eagle}). These models project document images into a sequence of dense patch-level embeddings, enabling fine-grained, multi-vector matching against text queries. Additionally, current research in this space also actively exploring various optimization strategies to refine performance and efficiency. These include knowledge distillation from high-capacity teachers \cite{santhanam2022colbertv2},  quantization for reduced memory footprints \cite{santhanam2022colbertv2}, and model merging \cite{qwen3-embed,gemini-embed,embedding-gemma,babakhin2025llama} to ensemble the weights of diverse pre-trained models.

Our work builds upon this emerging paradigm of VLM-based late interaction. We leverage pretrained VLM models as backbones for representation extraction and employ a late interaction mechanism specifically tailored for visually rich documents.

\subsection{Dense Retrieval} 

Dense retrieval methods for both text and visual documents are often categorized by how and when the query and document interact. Existing approaches follow three primary paradigms:

(1) \textbf{bi-encoders}: These models independently encode queries and documents into single global vectors. Relevance is determined by a simple similarity function, usually cosine similarity, as shown in Figure~\ref{fig:biencoder}. While highly efficient and common in early text-only retrieval \cite{e5} or CLIP/SigLIP-style VLMs \cite{clip, siglip}, the single-vector representation bottleneck is often limited to capture fine-grained lexical, visual, or layout-dependent cues.

(2) \textbf{Cross-encoders}: These architectures jointly encode queries and candidate documents, allowing for dense early interaction via self- and cross-attention over all tokens or patches. They can model intricate token–level alignments and thus excel as strong rerankers. However, it does not support pre-computing/indexing document embeddings, and the quadratic computational cost renders them impractical for first-stage retrievers across large-scale collections. 

(3) \textbf{Late interaction models}: Bridging the gap between the two, late interaction retains some expressiveness from cross-encoders, but still allows pre-computing documents' multi-vector embeddings, as further discussed in the next section.

\subsection{Late Interaction}

\begin{figure}[ht]
  \centering
  \begin{subfigure}[t]{0.48\textwidth}
    \centering
    \includegraphics[width=\linewidth]{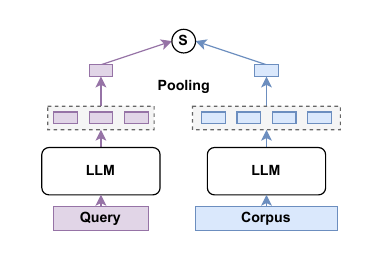}
    \caption{Bi-encoder architecture with Pooling}
    \label{fig:biencoder}
  \end{subfigure}
  \hfill
  \begin{subfigure}[t]{0.48\textwidth}
    \centering
    \includegraphics[width=\linewidth]{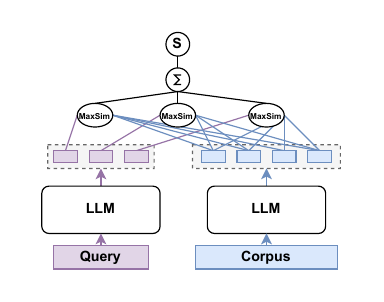}
    \caption{Late-interaction architecture}
  \label{fig:colbert}
  \end{subfigure}
  \caption{Illustration of the bi-encoder and late-interaction architectures.}
\end{figure}

The late interaction mechanism introduced by ColBERT~\cite{colbert} enables fine-grained interactions between query and document tokens. As shown in Figure~\ref{fig:colbert}, for a query, each token embedding interacts with all document token embeddings using a \emph{MaxSim} operator, which selects the maximum similarity per query token and sums these scores to produce the final relevance score.

This requires storing all token embeddings of the document corpus (text or images).
At inference time, query token embeddings are computed and interact with the stored document embeddings through \emph{MaxSim} op. We adopt this mechanism in our models to enable fine-grained retrieval.

While this approach offers the expressiveness of token-level matching, compared to simpler pooling methods such as average or last-token pooling, as shown in Figure~\ref{fig:biencoder}, the late-interaction method introduces latency and storage overhead that may need to be assessed, as these overheads become a concern for real-world applications. To ensure scalability, we further investigate dimensionality reduction techniques, refining the trade-off between dense representation quality and storage overhead, as discussed in Section~\ref{sec:tradeoff_challenges}.

\subsection{Contrastive Learning}

Dense retrieval models are typically trained with contrastive learning to maximize the embedding similarity between the query and positive passage, while minimizing the similarity between the query and negative corpus.
The InfoNCE contrastive loss~\cite{infonce} is a popular choice to train the model to distinguish between positive and negative pairs in a shared embedding space,

\begin{equation}
\mathcal{L}(q, d^+, D_N) = -\log \frac{\exp(\text{sim}(q, d^+)/\tau)}{\sum_{d_i \in \{d^+\} \cup D_N} \exp(\text{sim}(q, d_i)/\tau)},
\label{eq:infonce}
\end{equation}   

\noindent where $q$ is the embedding of a query, and $d^+$ are embeddings of positive documents. $D_N$ denotes the set of negative passages. $ \tau $ is the temperature parameter. $ sim(\cdot) $ represents a similarity function like cosine similarity, dot product, or late interaction similarity based on the \emph{MaxSim} operation.
\section{Nemotron ColEmbed V2: a Family of Multimodal Late Interaction Models for Visual Document Retrieval}

In this section, we describe the Nemotron ColEmbed V2 models' architectures and the main methods we used for those top performing models, listed in Table~\ref{tab:model_arch}.

\begin{table}[h]
  \centering 
  \caption{The Nemotron ColEmbed V2 family}
  \begin{threeparttable}

  \begin{tabular}{lcc}
    \toprule
     Model (Huggingface ID) & \# Parameters (without embeddings) & Embedding Dimension\\
    \midrule
     nvidia/llama-nemotron-colembed-3b-v2 & 4.40 (3.99) B & 3072 \\
     nvidia/nemotron-colembed-vl-4b-v2 &  4.82 (4.43) B & 2560 \\
     nvidia/nemotron-colembed-vl-8b-v2 &  8.76 (8.14) B & 4096 \\
    \bottomrule
  \end{tabular}  
  \end{threeparttable}
  \label{tab:model_arch}
\end{table}

\subsection{\emph{llama-nemotron-colembed-vl-3b-v2} Architecture}

Our \emph{llama-nemotron-colembed-vl-3b-v2} late-interaction VLM embedding model is based on NVIDIA Eagle 2 vision-language model~\cite{eagle, eagle25}. It uses the same architecture as its first version \emph{llama-nemoretriever-colembed-3b-v1} \footnote{The \emph{llama-nemotron-colembed-vl-3b-v2} is trained with 50\% more data than its former version (v1), including 115k augmented queries translated to different languages.}\cite{xu2025llama}. These models adopt dynamic image tiling to support inputs of varying resolutions, and employ a carefully curated data strategy that improves multimodal learning. These design choices enable Eagle 2 models to achieve state-of-the-art results on several multimodal benchmarks, providing a solid foundation for retrieval tasks. We initialize our model from an internal pre-trained Eagle 2 VLM that uses SigLip 2 \cite{siglip} as the image encoder and Llama 3.2 3B \cite{dubey2024llama} as the LLM backbone.

Regarding the dynamic tiling mechanism, the \texttt{max\_input\_tiles} parameter is used to control the number of tiles produced from each image. Each image tile generates 256 visual tokens. For training, we set \texttt{max\_input\_tiles = 2} (including an additional thumbnail tile from the full page) to maintain memory efficiency, as increasing it to 4 did not yield performance gains. During inference, we set \texttt{max\_input\_tiles = 8} to allow for finer visual granularity. 

\begin{figure}[ht]
  \centering
  \includegraphics[width=\linewidth]{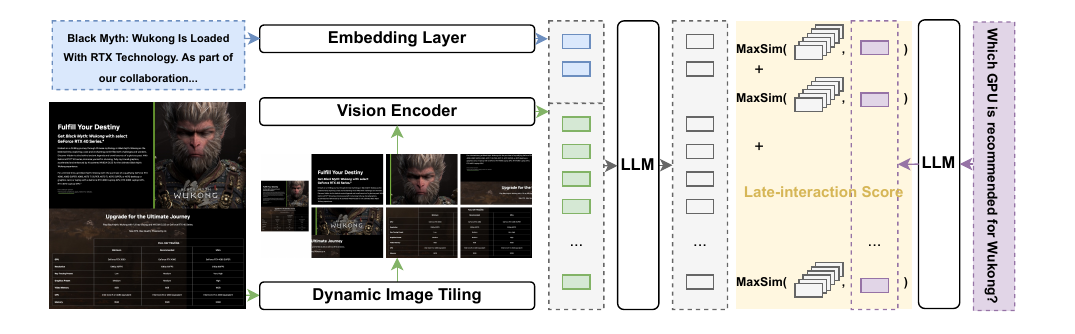}
  \caption{\emph{llama-nemotron-colembed-vl-3b-v2} architecture with dynamic image tiling and late interaction scoring mechanisms\cite{xu2025llama}.}
  \label{fig:colembed_3b}
\end{figure}

Figure~\ref{fig:colembed_3b} illustrates the dynamic image tiling, image encoding into visual tokens, and the late interaction mechanism.

\subsection{\emph{nemotron-colembed-vl-4/8b-v2} Architectures}

Our \emph{nemotron-colembed-vl-4b-v2} and \emph{nemotron-colembed-vl-8b-v2} late-interaction embedding models are based on Qwen3-VL 4B and 8B VLMs\cite{bai2025qwen3vltechnicalreport}, which support multimodal inputs and long context.
Similarly to Eagle 2 used by \emph{llama-nemotron-colembed-vl-3b-v2}, Qwen3-VL adopts a three-module architecture, comprising SigLIP-2 vision encoder\cite{siglip}, a two-layer MLP-based vision–language merger, and Qwen3 LLMs \cite{yang2025qwen3}. 

Their vision encoder is designed to handle dynamic, native-resolution images, mapping them to visual token sequences of variable length. To enhance perceptual capability and preserve rich visual details, Qwen3-VL extends the DeepStack mechanism by injecting  visual tokens from intermediate layers of the vision encoder into multiple layers of the LLM \cite{bai2025qwen3vltechnicalreport}.

\subsection{Training data and Hyperparameters}
The ColEmbed v2 models were trained on 500,000 samples sampled from publicly available datasets, including Vidore-ColPali-Training \cite{colpali}, Wiki-SS-NQ and DocMatix-IR \cite{ma2024unifying}, VDR\footnote{https://huggingface.co/datasets/llamaindex/vdr-multilingual-train}, and VisRAG \cite{yu2024visrag}. We employed some data-related techniques (hard-negative mining, cluster-based data sampling, cross-lingual query translation, and two-stage training) that are described in Section~\ref{sec:key_methods}.

ColEmbed v2 models were trained for one epoch, with a learning rate of 2e-6, AdamW optimizer, and weight decay of 0.1, batch size and gradient accumulation of 1, and two negatives per sample. The ColEmbed 8B v2 training takes 3h40m\footnote{We experimented with LoRA and could get a slightly lower accuracy, while training time reduces from 3h40m to 1h40m.} using 8x A100 80GB GPU cards.

\subsection{Key Methods for Better Performance}
\label{sec:key_methods}

\subsubsection{Modifying LLM decoder causal attention to bi-direction attention for encoders}

LLMs and VLMs are decoder models and use causal (uni-directional) attention. It means during training, when predicting a token, the model is prevented from accessing the following tokens (on the right) to avoid leaking information that would not be available during inference.

When adapting decoder LLMs as embedding models (encoders), a common practice involves transitioning the uni-directional attention to bi-directional attention. This modification enables Transformer layers to attend to the full context of a sequence, allowing each token to integrate information from both preceding and succeeding tokens. Such global representation has been shown to significantly enhance retrieval accuracy for LLM-based embedding models \cite{lee2024nv,nv-retriever,moreira2024enhancing}. Consistent with these findings, we implement this technique in Nemotron ColEmbed V2 models, where we observed substantial performance improvement when adapting Eagle 2 and Qwen3-VL architectures.

\subsubsection{Hard-Negative Mining}

Embedding models for retrieval are primarily trained with contrastive learning, a paradigm requiring triplets composed of a query, positive examples, and negative examples. Extensive research on information retrieval indicates the efficiency of contrastive learning hinges on the availability of hard-negatives, i.e., false examples that exhibit high semantic or lexical similarity to the query. The hard-negatives can be mined from the corpus using external sparse or dense "teacher"  retrieval models. 

For Nemotron ColEmbed V2 models, we used an internal Llama-Eagle 3B VLM embedding model for mining from the corpus the top-k most similar page images to the queries.

We leverage the \textit{top-k with percentage to positive threshold} method from NV-Retriever~\cite{nv-retriever}. It filters the potential hard-negatives by limiting their maximum similarity scores to a percentage of the positive sample similarity score, creating a margin that reduces the number of false negatives (e.g. the ones that should be actually positives). We set the threshold as 0.95, meaning we select the $K$ most relevant negative samples whose similarity to the query is less than 95\% of the query–positive similarity score.

This encourages the model to learn from challenging negatives, while removing potential false negatives that have high similarity scores.

\subsubsection{Cluster-based Data Sampling}

Public training datasets typically exhibit imbalance, with disproportionate numbers of samples across different domains. Training on such skewed data could lead to overfitting to specific domains, which compromises the model's ability to generalize to others, especially the underrepresented ones.

To mitigate this, Nemotron-CLIMB\cite{diaonemotron} proposed a framework for optimizing the blend of LLM training data by partitioning the corpus into distinct clusters and sampling a curated percentage of training examples from each. 

For Nemotron ColEmbed V2 models, we adapted their method, clustering the positive contexts, then sampling some positive samples from each cluster together with the associated queries and negatives. 

For clustering, we generate embeddings from document page images using an internal Llama-Eagle VLM embedding model, resulting in 3072-dim vectors. We then apply PCA to reduce embeddings dimension to 50, followed by a K-Means clustering approach utilizing gap statistics \cite{tibshirani2001estimating} to choose the most representative \emph{k} clusters.
To ensure diversity and balance the domains of our training blend, we perform uniform sampling from the 14 discovered clusters.

\subsubsection{Cross-lingual Translation}

There has been growing interest from the community in multi-lingual and cross-lingual retrieval, the latter for cases where the query and corpus languages are different. This is particularly challenging for visual document retrieval, as document pages are represented as images to the model.
Vidore V3 exemplifies this interest, as the corpus of PDF documents is in English or French, and queries are translated into six languages:  English, French, Spanish, German, Italian, and Portuguese.

To enable Nemotron ColEmbed V2 models to support cross-lingual retrieval, 
we have augmented our training data using \emph{Qwen3-235B-A22} model to translate sampled queries from each discovered cluster into other languages.

\subsubsection{Two-stage Training}

The \emph{llama-nemotron-colembed-vl-3b-v2} model is trained in two stages \cite{xu2025llama}. In the first stage, the model is trained on a textual corpus consisting of query-positives-negatives triplets. This stage is designed to establish a robust foundation for semantic similarity within the textual embedding space.

In the second stage, we fine-tune the model on an image-retrieval corpus. This stage facilitates cross-modal alignment by grounding visual features in the textual representation space.

For \emph{nemotron-colembed-vl-4b-v2} and \emph{nemotron-colembed-vl-8b-v2}, due to Qwen3-VL's strong cross-modal pre-training, we perform a single-stage contrastive learning training with image corpus.

\subsubsection{Model Merging}

Model merging or model souping is a technique that combines weights of multiple models, typically those sharing the same architecture but trained with different data or hyperparameters \cite{model-merge-2018, model-merge-2022}. It is observed that the efficacy of this approach increases with the diversity of the constituent models' weights. Recently, this approach has been popularized for improving the robustness and generalization of embedding models, such as Qwen3-Embedding~\cite{qwen3-embed}, Gemini~Embedding~\cite{gemini-embed},  EmbeddingGemma~\cite{embedding-gemma}, and Llama-Embed-Nemotron-8B~\cite{babakhin2025llama}.

For Nemotron ColEmbed V2 models, we employed a simple weighted average of model weights for merging. The 3B model is an ensemble of 8 individual models, and the 4B and 8B are merged from 4 individual models each. The individual models are trained with variations in the training blend (datasets and data size) and in the random seed that is used by the dataloader for sampling.

By comparing the ensembled model with the best individual model used in the ensemble, we have noticed that the accuracy gains might scale with the model size. We observed an improvement of 0.8\% for \emph{llama-nemotron-colembed-vl-3b-v2}, 1.0\% for the \emph{nemotron-colembed-vl-4b-v2}, and 1.5\% for the \emph{nemotron-colembed-vl-8b-v2} model.
\section{Results}

In this section we demonstrate the effectiveness of Nemotron ColEmbed V2 models on different visual document retrieval benchmarks, with \emph{nemotron-colembed-vl-8b-v2} being the top-performer in all of them.

\subsection{Vidore V3 Leaderboard}

The Vidore V3~\cite{loison2026vidorev3} is a recent benchmark that emphasizes evaluation of vision document retrieval for complex enterprise and real-world scenarios, including multi-type and multi-language queries across ten professional domains. 

To ensure submitted models do not overfit to the leaderboard, Vidore V3 has created two sets of tasks/datasets: eight public and two private, for which test datasets were not released publicly. The MTEB maintainers have established a process in which they run themselves the submitted models evaluation on the private tasks and report in the leaderboard.

%To privilege models that perform well in most of the tasks and not overfit to a few tasks, the MTEB ViDoRe leaderboards sorts models not by Avg. NDCG, but rather by Borda rank. That method starts, for each task, by sorting  models per task NDCG in ascending order, and then giving each model a rank score for the task, so that the model with the higher NDCG gets the highest rank for the task. All model ranks per task are then summed (Borda count). Then models are sorted by their Borda count and the Borda rank is their leaderboard position.

Table~\ref{tab:vidorev3} presents the retrieval accuracy (NDCG@10) of the top  models in MTEB Vidore V3 leaderboard, as of Feb. 03, 2026. We can see that our \emph{nemotron-colembed-vl-8b-v2} model places 1st in the leaderboard, with NDCG@10 of 63.42, with +3\% improvement to the second place. The leaderboard also presents results for  \emph{nemotron-colembed-vl-4b-v2} and
 \emph{llama-nemotron-colembed-vl-3b-v2}, which demonstrate the highest Avg. NDCG@10 over the other models with the same size.

\begin{table}[h]
\caption{Vidore V3 leaderboard as of Feb. 03, 2026. Our Nemotron ColEmbed V2 models are highlighted in gray. Retrieval accuracy scores are NDCG@10.}

\centering
  \resizebox{\textwidth}{!}{%
  
\begin{tabular}{clc|cccccccc|cc}
 & & & \multicolumn{8}{c}{\textbf{Public tasks}} & \multicolumn{2}{c}{\textbf{Private tasks}} \\
\textbf{\makecell[l]{Rank}} &
\textbf{Model}                     & \textbf{Avg}   & \textbf{CompSci} & \textbf{Energy} & \textbf{FinanceEn} & \textbf{FinanceFr} & \textbf{HR} & \textbf{Industrial} & \textbf{Pharma} & \textbf{Physics} & \textbf{Nuclear} & \textbf{Telecom} \\ \toprule
\rowcolor[HTML]{EFEFEF} 
1 & nemotron-colembed-vl-8b-v2               & \textbf{63.42} & 79.30            & 69.82           & 67.29              & 51.54              & 66.32       & 56.03               & 67.19           & 50.84    &  53.84  &  72.00       \\
2 & tomoro-colqwen3-embed-8b           & \textbf{61.59} & 75.35            & 68.41           & 65.08              & 49.10              & 63.98       & 54.41               & 66.36           & 50.13       &  52.65  &  70.46        \\
\rowcolor[HTML]{EFEFEF} 
3 & nemotron-colembed-vl-4b-v2               & \textbf{61.54} & 78.56            & 67.48           & 65.02              & 49.01              & 62.39       & 53.91               & 66.10           & 48.86         &  52.78  &     71.30   \\
4 & Ops-Colqwen3-4B               & \textbf{61.17} & 77.74            & 66.49           & 65.71              & 48.81              & 61.81       & 53.99               & 66.42           & 49.14      &  52.23  &  69.33         \\
5 & tomoro-colqwen3-embed-4b           & \textbf{60.20} & 75.44            & 66.43           & 63.84              & 46.83              & 60.09       & 53.58               & 65.74           & 49.32        &  51.23  &  69.44       \\
\rowcolor[HTML]{EFEFEF} 
6 & llama-nemotron-colembed-vl-3b-v2            & \textbf{59.79} & 77.09            & 64.88           & 64.23              & 44.41              & 62.28       & 51.71               & 66.04           & 46.93       &  50.65  &  69.68        \\

7 & jina-embeddings-v4                 & \textbf{57.52} & 71.81            & 63.50           & 59.30              & 46.10              & 59.53       & 50.38               & 63.09           & 46.63       &  50.02  &  64.81        \\

8 & colnomic-embed-multimodal-7b       & \textbf{57.33} & 76.20            & 63.58           & 56.57              & 45.46              & 58.67       & 50.13               & 62.26           & 48.25       &  45.02  &   67.16       \\

9 & llama-nemoretriever-colembed-3b-v1 & \textbf{57.26} & 75.16            & 62.07           & 60.88              & 43.77              & 58.69       & 47.09               & 63.74           & 45.13     &  49.15  & 64.74  \\

%9 & SauerkrautLM-ColQwen3-8b-v0.1       & \textbf{58.54} & 77.52            & 66.32           & 55.79              & 45.03              & 59.96       & 50.38               & 63.97           & 49.35        &    &         \\

       \bottomrule 
       
\end{tabular}
}

\label{tab:vidorev3}
\end{table}

\subsection{Vidore V1\&V2 Leaderboard}

Table~\ref{tab:vidore_v1_v2} presents results for the MTEB Vidore V1\&V2 leaderboard as of Feb. 03, 2026.

The leaderboard integrates the two benchmarks because Vidore V1 provided public in-domain training data, and many models overfit to it. For Vidore V2, no in-domain training data was provided, as well for Vidore V3.

Our \emph{nemotron-colembed-vl-8b-v2} model places second, close to the leading model. The \emph{llama-nemotron-colembed-vl-3b-v2} and \emph{nemotron-colembed-vl-4b-v2} are also among the top-4.

% Please add the following required packages to your document preamble:
% \usepackage[table,xcdraw]{xcolor}
% Beamer presentation requires \usepackage{colortbl} instead of \usepackage[table,xcdraw]{xcolor}
\begin{table}[ht]
\caption{Vidore V1\&V2 leaderboard as of Feb. 03, 2026. Our Nemotron ColEmbed V2 models are highlighted in gray. Retrieval accuracy scores are NDCG@5.}
\centering
  \resizebox{\textwidth}{!}{%

\begin{tabular}{clc|cccccccccc|cccc}
%\multicolumn{1}{l}{}                    &                                    & \multicolumn{1}{l}{}              & \multicolumn{10}{c}{\textbf{Vidore V1}}                                                                                                                                                                                                                                                                                                                                                                     & \multicolumn{4}{c}{\textbf{Vidore V2}}                                                                                                                                             \\
%\multicolumn{1}{l}{\textbf{\makecell[l]{Borda\\rank}}} & \textbf{Model}                     & \multicolumn{1}{l}{\textbf{Avg.}} & \multicolumn{1}{l}{\textbf{ArxivQA}} & \multicolumn{1}{l}{\textbf{DocVQA}} & \multicolumn{1}{l}{\textbf{InfoVQA}} & \multicolumn{1}{l}{\textbf{\makecell[l]{Shift\\Project}}} & \multicolumn{1}{l}{\textbf{AI}} & \multicolumn{1}{l}{\textbf{Energy}} & \multicolumn{1}{l}{\textbf{\makecell[l]{Gov.\\Reports}}} & \multicolumn{1}{l}{\textbf{Healthcare}} & \multicolumn{1}{l}{\textbf{TabFQuad}} & \multicolumn{1}{l}{\textbf{TAT-DQA}} & \multicolumn{1}{l}{\textbf{\makecell[l]{MIT\\Biomed.}}} & \multicolumn{1}{l}{\textbf{\makecell[l]{ESG\\Restau.\\En}}} & \multicolumn{1}{l}{\textbf{\makecell[l]{ESG\\Restau.\\Multi}}} & \multicolumn{1}{l}{\textbf{\makecell[l]{Econ.\\Macro}}} \\ \toprule
\multirow{2}{*}{\textbf{Rank}} & \multicolumn{1}{c}{\multirow{2}{*}{\textbf{Model}}} & \multirow{2}{*}{\textbf{Avg.}} & \multicolumn{10}{c}{\textbf{Vidore V1}} & \multicolumn{4}{c}{\textbf{Vidore V2}} \\ 
\cmidrule(lr){4-13} \cmidrule(lr){14-17}
& & & ArxivQA & DocVQA & InfoVQA & Shift Project & AI & Energy & Gov. Reports & Healthcare & TabFQuad & TAT-DQA & MIT Biomed. & ESG Restau. En & ESG Restau. Multi & Econ. Macro \\ 
\midrule
\textbf{1}                              & Ops-Colqwen3-4B                    & \textbf{84.87}                             & 91.78                                & 66.45                               & 94.02                                & 90.84                                      & 99.63                           & 97.26                               & 98.02                                     & 99.63                                   & 93.55                                 & 82.38                                & 65.53                                    & 78.61                                       & 66.05                                          & 64.45                                    \\
\rowcolor[HTML]{EFEFEF} 
\textbf{2}                              & nemotron-colembed-vl-8b-v2         & \textbf{84.80}                             & 93.08                                & 68.05                               & 94.56                                & 93.30                                      & 100.00                          & 97.89                               & 98.89                                     & 99.63                                   & 97.74                                 & 83.37                                & 66.16                                    & 73.15                                       & 60.56                                          & 60.76                                    \\
\rowcolor[HTML]{EFEFEF} 
\textbf{3}                              & nemotron-colembed-vl-4b-v2         & \textbf{83.87}                             & 92.03                                & 67.39                               & 93.31                                & 92.26                                      & 99.26                           & 96.19                               & 98.02                                     & 98.52                                   & 98.05                                 & 81.19                                & 64.32                                    & 71.43                                       & 61.48                                          & 60.75                                    \\
\rowcolor[HTML]{EFEFEF} 
\textbf{4}                              & llama-nemotron-colembed-vl-3b-v2   & \textbf{83.64}                             & 90.40                                & 67.17                               & 94.68                                & 92.00                                      & 100.00                          & 98.02                               & 97.95                                     & 98.89                                   & 97.25                                 & 81.04                                & 63.19                                    & 73.11                                       & 58.64                                          & 58.59                                    \\
\textbf{5}                              & tomoro-colqwen3-embed-8b           & \textbf{83.52}                             & 91.15                                & 66.37                               & 94.48                                & 87.89                                      & 99.26                           & 96.71                               & 97.58                                     & 99.06                                   & 94.23                                 & 80.92                                & 65.47                                    & 75.98                                       & 60.71                                          & 59.46                                    \\
\textbf{6}                              & EvoQwen2.5-VL-Retriever-7B-v1      & \textbf{83.41}                             & 91.49                                & 65.07                               & 94.11                                & 88.80                                      & 99.63                           & 96.63                               & 96.29                                     & 98.89                                   & 93.63                                 & 82.26                                & 65.20                                    & 76.98                                       & 59.67                                          & 59.13                                    \\
\textbf{7}                              & tomoro-colqwen3-embed-4b           & \textbf{83.18}                             & 90.58                                & 66.30                               & 94.31                                & 87.39                                      & 99.26                           & 96.91                               & 97.17                                     & 99.63                                   & 94.33                                 & 79.87                                & 65.38                                    & 74.65                                       & 62.44                                          & 56.30                                    \\
\textbf{8}                              & llama-nemoretriever-colembed-3b-v1 & \textbf{83.10}                             & 88.35                                & 66.21                               & 94.92                                & 90.70                                      & 99.63                           & 96.63                               & 97.82                                     & 99.26                                   & 95.94                                 & 80.57                                & 62.70                                    & 75.38                                       & 57.38                                          & 57.84                                    \\
\textbf{9}                             & SauerkrautLM-ColQwen3-8b-v0.1      & \textbf{82.91}                             & 93.80                                & 64.69                               & 94.51                                & 90.41                                      & 98.65                           & 96.52                               & 96.79                                     & 99.26                                   & 92.18                                 & 84.04                                & 63.26                                    & 70.77                                       & 57.85                                          & 57.98   \\  
\textbf{10}                              & EvoQwen2.5-VL-Retriever-3B-v1      & \textbf{82.76}                             & 90.46                                & 63.67                               & 92.22                                & 88.60                                      & 100.00                          & 97.63                               & 98.89                                     & 99.26                                   & 93.99                                 & 82.00                                & 63.63                                    & 67.11                                       & 59.05                                          & 62.19                                    \\

                 \bottomrule  

\end{tabular}
}

\label{tab:vidore_v1_v2}

\end{table}

\subsection{MIRACL-Vision benchmark}

MIRACL-Vision is a large multi-lingual VDR benchmark \cite{miracl-vision}, covering many popular and under-resourced languages. It is based on MIRACL\cite{miracl} text multilingual benchmark and the corpus is composed by Wikipedia passages. MIRACL-VISION corpus, instead, is composed of screenshot images extracted from Wikipedia pages.

Table~\ref{tab:miracl_vision} shows retrieval accuracy (NDCG@10) on MIRACL-Vision multilingual benchmark. It can be observed how models' accuracy on visual document retrieval vary for popular vs. under-resourced languages (e.g. Telugu, Yoruba).

Nemotron ColEmbed V2 models perform 
better among the analyzed models, because of the pre-training of the backbone LLMs and also our augmented training blend containing cross-lingual examples. The \emph{nemotron-colembed-vl-8b-v2} provides the highest score for most of the languages.

\begin{table}[ht]
\caption{MIRACL-Vision results (NDCG@10) on multi-lingual visual document retrieval. Nemotron ColEmbed V2 models are in the last three columns, results from other models obtained from MIRACL-Vision paper\cite{miracl-vision}.}

\centering
  \resizebox{\textwidth}{!}{%

\begin{tabular}{l|p{1.4cm}p{1.4cm}p{1.1cm}p{1.3cm}p{1.8cm}|p{1.6cm}p{1.5cm}p{1.5cm}}
\textbf{Language} & \emph{dse-qwen2-2b-mrl-v1} & \emph{gme-Qwen2-VL-2B-Instruct} & \emph{vdr-2b-multi-v1} & \emph{colqwen2-v1.0} & \emph{llama- nemoretriever-colembed-3b-v1} & \emph{nemoretriever-colembed-3b-v2} & \emph{nemotron-colembed-vl-4b-v2} & \emph{nemotron-colembed-vl-8b-v2} \\ \toprule
Arabic            & 0.3893                                                                   & 0.4888                                                                        & 0.4379                                                               & 0.4129                                                             & 0.4247                                                                                                                              & 0.5250                                                                            & 0.6028                                                                          & \textbf{0.7863}                                                                 \\
Bengali           & 0.2352                                                                   & 0.3755                                                                        & 0.2473                                                               & 0.2888                                                             & 0.4878                                                                                                                              & 0.5391                                                                            & 0.5156                                                                          & \textbf{0.6160}                                                                 \\
Chinese           & 0.5962                                                                   & 0.6314                                                                        & 0.5963                                                               & 0.4926                                                             & 0.4355                                                                                                                              & 0.4878                                                                            & 0.6697                                                                          & \textbf{0.7204}                                                                 \\
English           & 0.6605                                                                   & 0.6784                                                                        & 0.6784                                                               & 0.6417                                                             & 0.7363                                                                                                                              & 0.7397                                                                            & 0.7246                                                                          & \textbf{0.7480}                                                                 \\
Farsi             & 0.2250                                                                   & 0.3085                                                                        & 0.2398                                                               & 0.2616                                                             & 0.3109                                                                                                                              & 0.3570                                                                            & 0.4266                                                                          & \textbf{0.5289}                                                                 \\
Finnish           & 0.4162                                                                   & 0.6863                                                                        & 0.5283                                                               & 0.6604                                                             & 0.8513                                                                                                                              & 0.8541                                                                            & 0.8398                                                                          & \textbf{0.8726}                                                                 \\
French            & 0.7160                                                                   & 0.6851                                                                        & 0.7194                                                               & 0.6876                                                             & 0.7988                                                                                                                              & 0.7943                                                                            & 0.7943                                                                          & \textbf{0.8171}                                                                 \\
German            & 0.6267                                                                   & 0.6345                                                                        & 0.6205                                                               & 0.5995                                                             & 0.6831                                                                                                                              & 0.6924                                                                            & 0.7100                                                                          & \textbf{0.7233}                                                                 \\
Hindi             & 0.1740                                                                   & 0.3127                                                                        & 0.2058                                                               & 0.2209                                                             & 0.4867                                                                                                                              & 0.5319                                                                            & 0.5338                                                                          & \textbf{0.5902}                                                                 \\
Indonesian        & 0.4866                                                                   & 0.5416                                                                        & 0.5254                                                               & 0.5320                                                             & 0.6428                                                                                                                              & 0.6550                                                                            & 0.6480                                                                          & \textbf{0.6680}                                                                 \\
Japanese          & 0.6232                                                                   & 0.7305                                                                        & 0.6553                                                               & 0.6970                                                             & 0.7260                                                                                                                              & 0.7493                                                                            & 0.8326                                                                          & \textbf{0.8690}                                                                 \\
Korean            & 0.4446                                                                   & 0.6202                                                                        & 0.4952                                                               & 0.4419                                                             & 0.5158                                                                                                                              & 0.5394                                                                            & 0.6136                                                                          & \textbf{0.7316}                                                                 \\
Russian           & 0.6505                                                                   & 0.7202                                                                        & 0.6995                                                               & 0.6811                                                             & 0.7670                                                                                                                              & 0.7920                                                                            & 0.7879                                                                          & \textbf{0.8399}                                                                 \\
Spanish           & 0.5927                                                                   & 0.6277                                                                        & 0.6274                                                               & 0.6224                                                             & 0.7109                                                                                                                              & \textbf{0.7236}                                                                   & 0.7033                                                                          & 0.7089                                                                          \\
Swahili           & 0.4156                                                                   & 0.5348                                                                        & 0.4509                                                               & 0.4931                                                             & \textbf{0.7767}                                                                                                                     & 0.7495                                                                            & 0.6886                                                                          & 0.7422                                                                          \\
Telugu            & 0.0274                                                                   & 0.0893                                                                        & 0.0318                                                               & 0.0264                                                             & 0.1669                                                                                                                              & \textbf{0.2325}                                                                   & 0.1579                                                                          & 0.1899                                                                          \\
Thai              & 0.2692                                                                   & 0.3563                                                                        & 0.3177                                                               & 0.2389                                                             & 0.4035                                                                                                                              & 0.4727                                                                            & 0.5928                                                                          & \textbf{0.6699}                                                                 \\
Yoruba            & 0.4178                                                                   & 0.4884                                                                        & 0.4577                                                               & 0.5120                                                             & 0.5888                                                                                                                              & \textbf{0.5943}                                                                   & 0.4469                                                                          & 0.5252                                                                          \\ \midrule
\textbf{Average}  & \textbf{0.4426}                                                          & \textbf{0.5283}                                                               & \textbf{0.4741}                                                      & \textbf{0.4728}                                                    & \textbf{0.5841}                                                                                                                     & \textbf{0.6127}                                                                   & \textbf{0.6272}                                                                 & \textbf{0.6860}  \\
\bottomrule
\end{tabular}
}

\label{tab:miracl_vision}

\end{table}
\newpage
\section{Late-interaction Deployment Challenges}
\label{sec:tradeoff_challenges}

Leaderboards and benchmarks typically evaluate performance based on accuracy metrics. Some also include proxy indicators of computational efficiency, such as model size or embedding dimensionality. Ultimately, rankings are determined by accuracy, which may not reflect the broader needs of real-world applications. No solution fits all use-cases. In this section, we discuss trade-offs of late interaction in the context of production deployment.

\subsection{Accuracy Considerations}
To help understand the retrieval accuracy difference obtained with single vector (average pooling) vs late interaction, we have run experiments comparing both methods, using the same model architecture, training data and hyperparameters.
The results are reported in Table~\ref{tab:late_interaction_vs_single_vector}. We can see that the accuracy obtained with late interaction is about 12\% and 9\% higher for ColEmbed 4B and ColEmbed 8B, respectively. Such effectiveness is the reason why late interaction embedding models have dominated Visual Document Retrieval benchmarks like ViDoRe.

\begin{table}[ht]

\caption{Comparison retrieval accuracy (Vidore V3 NDCG@10) of late-interaction vs single vector (avg. pooling), using same training data and hyperparameters. }
\centering
  \resizebox{\textwidth}{!}{%\
\begin{tabular}{m{5.4cm}|cc|c}
\textbf{Model}          & \textbf{Single vector} & \textbf{Late interaction} & \textbf{\% improvement}  \\ \toprule
nemotron-colembed-vl-4b-v2             & 54.07             & 60.52  &   + 11.92\%        \\
nemotron-colembed-vl-8b-v2             & 56.91             & 62.24   &   + 9.36\%       \\
\bottomrule
\end{tabular}
}
\label{tab:late_interaction_vs_single_vector}
\end{table}

\subsection{Retrieval Systems Considerations}

Deploying a production system involves balancing accuracy, latency/throughput, and cost. Typical retrieval system requirements involve the following aspects:

\begin{itemize}
  \item \textbf{Model size:} All embeddings of documents are generated by retrieval model. This step can be performed in batches, with support for continuous updates as new documents arrive. Throughput and cost are key considerations, and the overall retrieval performance is primarily influenced by the size of the model.
  \item \textbf{Storage:} The embeddings storage requirements are primarily determined by the embedding dimension, numeric precision and number of vectors per document.
  \item \textbf{Serving:} Latency measures how quickly documents can be retrieved in response to a user query. Since queries are typically short (around 50–100 tokens), the size of the embedding model plays a smaller role in this stage. Incorporating a reranker in the retrieval pipeline, such as a cross-encoder, can improve accuracy, but at the cost of increasing the latency to serve another model.

\end{itemize}

We discuss the trade-offs between those aspects in the next section.

\subsection{Retrieval Pipelines Trade-offs}

The late-interaction paradigm \cite{colbert} has demonstrated significant performance improvements in retrieval tasks by preserving fine-grained token-level interactions between queries and documents. Unlike traditional pooling strategies that compress entire sequences into single vectors, late-interaction models leverage all token-level representations. However, this approach introduces a fundamental trade-off between accuracy and storage cost, as each document requires multiple token embeddings, leading to significantly increased storage requirements.

Table~\ref{tab:late_interaction_vs_reranker} summarizes these  trade-offs aspects for different retrieval approaches, reporting  requirements in GigaBytes (GB) for storing the embeddings for one million page images. The storage footprint of late-interaction models depends on three key factors: token/embedding count (sequence length), embedding dimension, and numerical precision (e.g., float32, float16, int8). As can be observed in the table, late interaction models require much more storage for indexing document multi-vector embeddings. 

\begin{table}[ht]

\caption{Comparison of different VLM models in terms of model size, embedding dimension, storage requirements and retrieval accuracy (Vidore V3 NDCG@10). Last line is a retrieval pipeline in which an embedding model retrieves the top 50 page images, subsequently reranked by a cross-encoder.}
\centering
  \resizebox{\textwidth}{!}{%\

\begin{tabular}{m{5.4cm}|cccccc}
\textbf{Model}                                                                  & \textbf{Params (B)} & \textbf{\makecell[l]{Embed.\\dim}}   & \textbf{\makecell[l]{Avg. tokens/\\embeddings \\per image}} & \textbf{\makecell[l]{\# floating\\points\\per image}} & \textbf{\makecell[l]{Storage for 1M\\fp16  embed.(GB)}} & \textbf{\makecell[l]{Vidore V3\\NDCG@10}} \\ \toprule

           nemotron-colembed-vl-8b-v2                                                      & 8.14                                     & 4096          & 773                                 & 3166208                      & 5897.5                                    & 63.42                              \\
nemotron-colembed-vl-4b-v2                                                      & 4.43                                     & 2560          & 773                                 & 1978880                      & 3686.0                                    & 61.42                              \\
llama-nemotron-colembed-vl-3b-v2                                                & 3.99                                     & 3072          & 2304                                & 7077888                      & 13183.6                                   & 59.70                              \\ \midrule
llama-nemoretriever-colembed-1b-v1                                  & 2.15                                     & 2048          & 2304                                & 4718592                      & 8789.1                                    & 55.48                              \\
llama-nemotron-embed-vl-1b-v2                                    & 1.41                                     & 2048          & 1                                   & 2048                         & 3.8                                       & 48.69                              \\
llama-nemotron-embed-vl-1b-v2 w/ llama-nemotron-rerank-vl-1b-v2 & 1.41 + 1.41                              & 2048          & 1                                   & 2048                         & 3.8                                       & 54.41                              \\
\bottomrule
\end{tabular}
}
\label{tab:late_interaction_vs_reranker}
\end{table}

The number of tokens per image (sequence length) is determined by the VLM image tiling/resizing logic and its image encoder. For example, Qwen-3 VL (backbone models from our 8B and 4B embedding models) generates on average 773 visual embeddings for Vidore V3 pages, while Eagle 2 generates 2304 visual embeddings for Vidore V3 page images (with resolution of about 1654x2339 pixels).

Late interaction models require orders of magnitude more storage. For instance, the \emph{llama-nemoretriever-colembed-1b-v1} late interaction model and the \emph{llama-nemotron-embed-vl-1b-v2} single-vector model. Both use Llama 3.2 1B as LLM backbone. While the single-vector model requires 3.8 GB for embedding storage, the late-interaction model requires 8,789.1 GB (2312x). If we add the \emph{llama-nemotron-rerank-vl-1b-v2} cross-encoder model to the retrieval pipeline, reranking the top-50 page images retrieved by \emph{llama-nemotron-embed-vl-1b-v2}, the NDCG@10 boosts from 48.69 to 54.40, which is close to the 55.48 accuracy from the late interaction \emph{llama-nemoretriever-colembed-1b-v1} for a small fraction of storage requirements\footnote{Although both models use Llama 3.2 1B as LLM backbone, this comparison requires some remarks as late interaction is not the only factor we change here: (1) those models are trained on different training blends, (2) \emph{llama-nemoretriever-colembed-1b-v1} uses a larger  image encoder (SigLIP2 1B) than \emph{llama-nemotron-embed-vl-1b-v2} (SigLIP2 400M).}.

Latency is another important challenge for Late-interaction models. During inference, it requires calculation between a query and the multi-vectors of all page images in the corpus. Late interaction requires specialized vector database support to the MaxSim operation, and might introduce latency overhead. The alternative pipeline composed of a single-vector embedding followed by a cross-encoder provides much lower latency, because the cross-encoder performs early-interaction between query and document tokens only for the top-k documents retrieved by the embedding model, and not for the whole corpus as in the late interaction approach.

Retrieval pipeline design that should be carefully aligned with the specific use case. For example, \cite{enhancingQAranker} results demonstrate that in scenarios where corpus is large and number of queries is moderate, a smaller less accurate embedding model combined with a reranker (to improve accuracy) can be more cost-efficient than a larger embedding model.

Ultimately, the choice between late-interaction and bi-encoder paradigms depends on specific use case requirements and system constraints.

\subsection{Ablation on Embedding-size Reduction}
Several techniques can minimize storage requirements for both paradigms, like reducing embedding dimensions. Linear projection layers can be used to downsize the embeddings output by the LLM backbone. Matryoshka Representation Learning~\cite{matryoshka} allows having a single model that outputs embeddings that can be sliced/pruned to multiple smaller dimensions.

Following the approach used in vidore/colqwen2-v1.0 models~\cite{colpali}, we applied a linear projection layer to reduce the output dimension to 512 and 128. To minimize accuracy variation in the ablation, for each architecture and dim size, we train four models with different seeds (that sample different portions of our data blend for training), and report the average NDCG@10 across these four models.

We can observe the ablation results in Table~\ref{tab:ablation_emb_size}. For the \emph{nemotron-colembed-vl-8b-v2}, projecting embeddings to 512-dim reduces storage requirements by 87.5\%, while keeping 96.02\% of retrieval accuracy. With 128-dim embeddings, it requires only 3\% of storage, keeping 95.36\% of the accuracy. We see a similar trend for \emph{nemotron-colembed-vl-4b-v2} model. However, even with 128-dim, the storage requirement of 184.3 GB for 1M pages may still be too high for production environments handling large document corpora.

\begin{table}[ht]

\caption{Ablation study on reducing the embedding sizes of late interaction models. For these models, we don't apply model merging; instead, we report the average NDCG@10 across four models trained with different seeds.}
\centering
  \resizebox{\textwidth}{!}{%\

\begin{tabular}{l|cccc|cc|cc}
\textbf{Model}                                                                  & \textbf{Params (B)} & \textbf{\makecell[l]{Embed.\\dim}}   & \textbf{\makecell[l]{Avg. tokens/\\embeddings \\per image}} & \textbf{\makecell[l]{\# floating\\points\\per image}} & \textbf{\makecell[l]{Storage for 1M\\fp16  embed.(GB)}} & \textbf{\% storage} & \textbf{\makecell[l]{Vidore V3\\NDCG@10}} & \textbf{\%  NDCG@10} \\ \toprule
\multirow{3}{*}{nemotron-colembed-vl-8b-v2} & \multirow{3}{*}{8.14}                    & 4096          & 773                                 & 3166208                      & 5897.5                                    & 100\%      & 62.29                              & 100.00\%      \\
                                &                                          & 512           & 773                                 & 395776                       & 737.2                                     & 13\%       & 59.81                              & 96.02\%       \\
                                &                                          & 128           & 773                                 & 98944                        & 184.3                                     & 3\%        & 59.40                              & 95.36\%       \\ \midrule
\multirow{3}{*}{nemotron-colembed-vl-4b-v2} & \multirow{3}{*}{4.43}                    & 2560          & 773                                 & 1978880                      & 3686.0                                    & 100\%      & 60.42                              & 100.00\%      \\
                                &                                          & 512           & 773                                 & 395776                       & 737.2                                     & 20\%       & 59.29                              & 98.13\%       \\
                                &                                          & 128           & 773                                 & 98944                        & 184.3                                     & 5\%        & 58.47                              & 96.77\%       \\
\bottomrule
\end{tabular}
}
\label{tab:ablation_emb_size}
\end{table}

Decreasing embeddings numerical precision, i.e., to float16 or int8, is an alternative to reducing storage footprint. Many vector databases already support storage and retrieval of lower-precision embeddings, some of them offering post-training quantization for precision reduction.

Additionally, binary quantization reduces precision to 1-bit per element, potentially reducing storage by 16x. However, our experience with bi-encoders indicates that binary quantization performs poorly when the embedding dimensionality is too small, and these techniques require further testing with late-interaction embedding size of 128. AnswerAI's late-pooling approach~\cite{answerai-latepooling} can reduce token vectors by factors of 3-5, while MUVERA~\cite{muvera} proposes converting multi-vector embeddings into single Fixed Dimensional Encodings (FDEs) whose inner product approximates multi-vector similarity, enabling the use of standard single-vector retrieval with smaller total embedding size.

%s\input{content/6_relatedworks}
\section{Conclusion}

In this paper, we introduce the Nemotron ColEmbed V2 family of late-interaction models for visual document retrieval. We demonstrate their top-performance on ViDoRe benchmarks and multi-lingual capabilities on MIRACL-Vision benchmark. We describe the main methods that boosted the accuracy of our late-interaction models, like changing VLMs backbones to use bi-directional attention, using positive-aware hard-negative mining, cluster-based data sampling, cross-lingual translation, and model merging. Finally, we discuss the deployment challenges of late-interactions models and highlight key considerations for real-world deployment. We present numbers that illustrate the trade-offs between accuracy and storage requirements and provide an ablation on reducing embedding sizes, thus storage requirements. 
Our release of Nemotron ColEmbed V2 late-interaction models provides a strong foundation for future research and practical applications in visual document retrieval.  

We recommend further research on reducing the storage requirements of late interaction models, e.g., by using smaller embedding dimensions or lower numerical precision, or fewer embeddings per sample, without significantly penalizing retrieval accuracy.

%% The acknowledgments section is defined using the "acknowledgments" environment
%% (and NOT an unnumbered section). This ensures the proper
%% identification of the section in the article metadata, and the
%% consistent spelling of the heading.
%\begin{acknowledgments}
%  ...  
%\end{acknowledgments}
%% The declaration on generative AI comes in effect
%% in Janary 2025. See also
%% https://ceur-ws.org/GenAI/Policy.html
%\section*{Declaration on Generative AI}
%The author(s) have not employed any Generative AI tools.
%The author(s) used Gemini in order to: Grammar and spelling check, Paraphrase and reword. The author(s) reviewed and edited the content as needed and take(s) full responsibility for the publication’s content.

%%
%% Define the bibliography file to be used
\bibliography{references}

@article{nv-embed,
  title={Nv-embed: Improved techniques for training llms as generalist embedding models},
  author={Lee, Chankyu and Roy, Rajarshi and Xu, Mengyao and Raiman, Jonathan and Shoeybi, Mohammad and Catanzaro, Bryan and Ping, Wei},
  journal={arXiv preprint arXiv:2405.17428},
  year={2024}
}

@article{nv-retriever,
  title={NV-Retriever: Improving text embedding models with effective hard-negative mining},
  author={Moreira, Gabriel de Souza P and Osmulski, Radek and Xu, Mengyao and Ak, Ronay and Schifferer, Benedikt and Oldridge, Even},
  journal={arXiv preprint arXiv:2407.15831},
  year={2024}
}

@article{qwen3-embed,
  title={Qwen3 Embedding: Advancing Text Embedding and Reranking Through Foundation Models},
  author={Zhang, Yanzhao and Li, Mingxin and Long, Dingkun and Zhang, Xin and Lin, Huan and Yang, Baosong and Xie, Pengjun and Yang, An and Liu, Dayiheng and Lin, Junyang and Huang, Fei and Zhou, Jingren},
  journal={arXiv preprint arXiv:2506.05176},
  year={2025}
}

@article{e5-mistral,
  title={Improving Text Embeddings with Large Language Models},
  author={Wang, Liang and Yang, Nan and Huang, Xiaolong and Yang, Linjun and Majumder, Rangan and Wei, Furu},
  journal={arXiv preprint arXiv:2401.00368},
  year={2023}
}

@article{mteb,
  title={MTEB: Massive text embedding benchmark},
  author={Muennighoff, Niklas and Tazi, Nouamane and Magne, Lo{\"\i}c and Reimers, Nils},
  journal={arXiv preprint arXiv:2210.07316},
  year={2022}
}

@inproceedings{clip,
  title={Learning transferable visual models from natural language supervision},
  author={Radford, Alec and Kim, Jong Wook and Hallacy, Chris and Ramesh, Aditya and Goh, Gabriel and Agarwal, Sandhini and Sastry, Girish and Askell, Amanda and Mishkin, Pamela and Clark, Jack and others},
  booktitle={International conference on machine learning},
  pages={8748--8763},
  year={2021},
  organization={PmLR}
}

@article{siglip,
  title={Siglip 2: Multilingual vision-language encoders with improved semantic understanding, localization, and dense features},
  author={Tschannen, Michael and Gritsenko, Alexey and Wang, Xiao and Naeem, Muhammad Ferjad and Alabdulmohsin, Ibrahim and Parthasarathy, Nikhil and Evans, Talfan and Beyer, Lucas and Xia, Ye and Mustafa, Basil and others},
  journal={arXiv preprint arXiv:2502.14786},
  year={2025}
}

@article{eagle,
  title={Eagle 2: Building Post-Training Data Strategies from Scratch for Frontier Vision-Language Models},
  author={Li, Zhiqi and Chen, Guo and Liu, Shilong and Wang, Shihao and VS, Vibashan and Ji, Yishen and Lan, Shiyi and Zhang, Hao and Zhao, Yilin and Radhakrishnan, Subhashree and others},
  journal={arXiv preprint arXiv:2501.14818},
  year={2025}
}

@article{e5,
  title={Text embeddings by weakly-supervised contrastive pre-training},
  author={Wang, Liang and Yang, Nan and Huang, Xiaolong and Jiao, Binxing and Yang, Linjun and Jiang, Daxin and Majumder, Rangan and Wei, Furu},
  journal={arXiv preprint arXiv:2212.03533},
  year={2022}
}

@inproceedings{colbert,
  title={Colbert: Efficient and effective passage search via contextualized late interaction over bert},
  author={Khattab, Omar and Zaharia, Matei},
  booktitle={Proceedings of the 43rd International ACM SIGIR conference on research and development in Information Retrieval},
  pages={39--48},
  year={2020}
}

@inproceedings{infonce,
  title={A simple framework for contrastive learning of visual representations},
  author={Chen, Ting and Kornblith, Simon and Norouzi, Mohammad and Hinton, Geoffrey},
  booktitle={International conference on machine learning},
  pages={1597--1607},
  year={2020},
  organization={PmLR}
}

@misc{colpali,
      title={ColPali: Efficient Document Retrieval with Vision Language Models}, 
      author={Manuel Faysse and Hugues Sibille and Tony Wu and Bilel Omrani and Gautier Viaud and Céline Hudelot and Pierre Colombo},
      year={2024},
      eprint={2407.01449},
      archivePrefix={arXiv},
      primaryClass={cs.IR},
      url={https://arxiv.org/abs/2407.01449}, 
}

@article{eagle25,
  title={Eagle 2.5: Boosting long-context post-training for frontier vision-language models},
  author={Chen, Guo and Li, Zhiqi and Wang, Shihao and Jiang, Jindong and Liu, Yicheng and Lu, Lidong and Huang, De-An and Byeon, Wonmin and Le, Matthieu and Rintamaki, Tuomas and others},
  journal={arXiv preprint arXiv:2504.15271},
  year={2025}
}

@inproceedings{enhancingQAranker,
  title={Enhancing Q\&A Text Retrieval with Ranking Models: Benchmarking, fine-tuning and deploying Rerankers for RAG},
  author={Moreira, Gabriel de Souza P and Ak, Ronay and Schifferer, Benedikt and Xu, Mengyao and Osmulski, Radek and Oldridge, Even},
  booktitle={Proceedings of the 1st Workshop on GenAI and RAG Systems for Enterprises, co-located with CIKM},
  year={2024}
}

@inproceedings{cradio,
  title={Am-radio: Agglomerative vision foundation model reduce all domains into one},
  author={Ranzinger, Mike and Heinrich, Greg and Kautz, Jan and Molchanov, Pavlo},
  booktitle={Proceedings of the IEEE/CVF Conference on Computer Vision and Pattern Recognition},
  pages={12490--12500},
  year={2024}
}

@article{vidore2,
  title={ViDoRe Benchmark V2: Raising the Bar for Visual Retrieval},
  author={Mac{\'e}, Quentin and Loison, Ant{\'o}nio and Faysse, Manuel},
  journal={arXiv preprint arXiv:2505.17166},
  year={2025}
}

@article{qwenvl,
  title={Qwen2-vl: Enhancing vision-language model's perception of the world at any resolution},
  author={Wang, Peng and Bai, Shuai and Tan, Sinan and Wang, Shijie and Fan, Zhihao and Bai, Jinze and Chen, Keqin and Liu, Xuejing and Wang, Jialin and Ge, Wenbin and others},
  journal={arXiv preprint arXiv:2409.12191},
  year={2024}
}

@article{nemotron,
  title={Llama-nemotron: Efficient reasoning models},
  author={Bercovich, Akhiad and Levy, Itay and Golan, Izik and Dabbah, Mohammad and El-Yaniv, Ran and Puny, Omri and Galil, Ido and Moshe, Zach and Ronen, Tomer and Nabwani, Najeeb and others},
  journal={arXiv preprint arXiv:2505.00949},
  year={2025}
}

@article{miracl,
  title={Miracl: A multilingual retrieval dataset covering 18 diverse languages},
  author={Zhang, Xinyu and Thakur, Nandan and Ogundepo, Odunayo and Kamalloo, Ehsan and Alfonso-Hermelo, David and Li, Xiaoguang and Liu, Qun and Rezagholizadeh, Mehdi and Lin, Jimmy},
  journal={Transactions of the Association for Computational Linguistics},
  volume={11},
  pages={1114--1131},
  year={2023},
  publisher={MIT Press One Broadway, 12th Floor, Cambridge, Massachusetts 02142, USA~…}
}

@article{mtebnew,
  title={Maintaining MTEB: Towards Long Term Usability and Reproducibility of Embedding Benchmarks},
  author={Chung, Isaac and Kerboua, Imene and Kardos, Marton and Solomatin, Roman and Enevoldsen, Kenneth},
  journal={arXiv preprint arXiv:2506.21182},
  year={2025}
}

@article{miracl-vision,
  title={MIRACL-VISION: A Large, multilingual, visual document retrieval benchmark},
  author={Osmulsk, Radek and Moreira, Gabriel de Souza P and Ak, Ronay and Xu, Mengyao and Schifferer, Benedikt and Oldridge, Even},
  journal={arXiv preprint arXiv:2505.11651},
  year={2025}
}

@online{answerai-latepooling,
  author = {Benjamin Clavié},
  title = {A little pooling goes a long way for multi-vector representations},
  year = 2024,
  url = {https://www.answer.ai/posts/colbert-pooling.html},
  urldate = {2015-07-02}
}

@article{muvera,
  title={MUVERA: multi-vector retrieval via fixed dimensional encodings},
  author={Dhulipala, Laxman and Hadian, Majid and Jayaram, Rajesh and Lee, Jason and Mirrokni, Vahab},
  journal={arXiv preprint arXiv:2405.19504},
  year={2024}
}

@article{matryoshka,
  title={Matryoshka representation learning},
  author={Kusupati, Aditya and Bhatt, Gantavya and Rege, Aniket and Wallingford, Matthew and Sinha, Aditya and Ramanujan, Vivek and Howard-Snyder, William and Chen, Kaifeng and Kakade, Sham and Jain, Prateek and others},
  journal={Advances in Neural Information Processing Systems},
  volume={35},
  pages={30233--30249},
  year={2022}
}

@article{paligemma2,
  title={Paligemma 2: A family of versatile vlms for transfer},
  author={Steiner, Andreas and Pinto, Andr{\'e} Susano and Tschannen, Michael and Keysers, Daniel and Wang, Xiao and Bitton, Yonatan and Gritsenko, Alexey and Minderer, Matthias and Sherbondy, Anthony and Long, Shangbang and others},
  journal={arXiv preprint arXiv:2412.03555},
  year={2024}
}

@inproceedings{xiao2024jina,
  title={Jina CLIP: Your CLIP model is also your text retriever},
  author={Xiao, Han and Mastrapas, Georgios and Wang, Bo},
  booktitle={Multi-modal Foundation Model meets Embodied AI Workshop@ ICML2024},
  year={2024}
}

@article{nussbaum2024nomic,
  title={Nomic embed vision: Expanding the latent space},
  author={Nussbaum, Zach and Duderstadt, Brandon and Mulyar, Andriy},
  journal={arXiv preprint arXiv:2406.18587},
  year={2024}
}

@article{loison2026vidorev3,
  title={ViDoRe V3: A Comprehensive Evaluation of Retrieval Augmented Generation in Complex Real-World Scenarios},
  author={Loison, Ant{\'o}nio and Mac{\'e}, Quentin and Edy, Antoine and Xing, Victor and Balough, Tom and Moreira, Gabriel and Liu, Bo and Faysse, Manuel and Hudelot, C{\'e}line and Viaud, Gautier},
  journal={arXiv preprint arXiv:2601.08620},
  year={2026}
}

@misc{bai2025qwen3vltechnicalreport,
      title={Qwen3-VL Technical Report}, 
      author={Shuai Bai and Yuxuan Cai and Ruizhe Chen and Keqin Chen and Xionghui Chen and Zesen Cheng and Lianghao Deng and Wei Ding and Chang Gao and Chunjiang Ge and Wenbin Ge and Zhifang Guo and Qidong Huang and Jie Huang and Fei Huang and Binyuan Hui and Shutong Jiang and Zhaohai Li and Mingsheng Li and Mei Li and Kaixin Li and Zicheng Lin and Junyang Lin and Xuejing Liu and Jiawei Liu and Chenglong Liu and Yang Liu and Dayiheng Liu and Shixuan Liu and Dunjie Lu and Ruilin Luo and Chenxu Lv and Rui Men and Lingchen Meng and Xuancheng Ren and Xingzhang Ren and Sibo Song and Yuchong Sun and Jun Tang and Jianhong Tu and Jianqiang Wan and Peng Wang and Pengfei Wang and Qiuyue Wang and Yuxuan Wang and Tianbao Xie and Yiheng Xu and Haiyang Xu and Jin Xu and Zhibo Yang and Mingkun Yang and Jianxin Yang and An Yang and Bowen Yu and Fei Zhang and Hang Zhang and Xi Zhang and Bo Zheng and Humen Zhong and Jingren Zhou and Fan Zhou and Jing Zhou and Yuanzhi Zhu and Ke Zhu},
      year={2025},
      eprint={2511.21631},
      archivePrefix={arXiv},
      primaryClass={cs.CV},
      url={https://arxiv.org/abs/2511.21631}, 
}

@article{beyer2024paligemma,
  title={Paligemma: A versatile 3b vlm for transfer},
  author={Beyer, Lucas and Steiner, Andreas and Pinto, Andr{\'e} Susano and Kolesnikov, Alexander and Wang, Xiao and Salz, Daniel and Neumann, Maxim and Alabdulmohsin, Ibrahim and Tschannen, Michael and Bugliarello, Emanuele and others},
  journal={arXiv preprint arXiv:2407.07726},
  year={2024}
}

@article{team2024qwen2,
  title={Qwen2 technical report},
  author={Team, Qwen and others},
  journal={arXiv preprint arXiv:2407.10671},
  volume={2},
  number={3},
  year={2024}
}

@inproceedings{gunther2025jina,
  title={jina-embeddings-v4: Universal embeddings for multimodal multilingual retrieval},
  author={G{\"u}nther, Michael and Sturua, Saba and Akram, Mohammad Kalim and Mohr, Isabelle and Ungureanu, Andrei and Wang, Bo and Eslami, Sedigheh and Martens, Scott and Werk, Maximilian and Wang, Nan and others},
  booktitle={Proceedings of the 5th Workshop on Multilingual Representation Learning (MRL 2025)},
  pages={531--550},
  year={2025}
}

@article{bai2025qwen25,
  title={Qwen2. 5-vl technical report},
  author={Bai, Shuai and Chen, Keqin and Liu, Xuejing and Wang, Jialin and Ge, Wenbin and Song, Sibo and Dang, Kai and Wang, Peng and Wang, Shijie and Tang, Jun and others},
  journal={arXiv preprint arXiv:2502.13923},
  year={2025}
}

@article{xu2025llama,
  title={Llama nemoretriever colembed: Top-performing text-image retrieval model},
  author={Xu, Mengyao and Moreira, Gabriel and Ak, Ronay and Osmulski, Radek and Babakhin, Yauhen and Yu, Zhiding and Schifferer, Benedikt and Oldridge, Even},
  journal={arXiv preprint arXiv:2507.05513},
  year={2025}
}

@inproceedings{xu2020layoutlm,
  title={Layoutlm: Pre-training of text and layout for document image understanding},
  author={Xu, Yiheng and Li, Minghao and Cui, Lei and Huang, Shaohan and Wei, Furu and Zhou, Ming},
  booktitle={Proceedings of the 26th ACM SIGKDD international conference on knowledge discovery \& data mining},
  pages={1192--1200},
  year={2020}
}

@inproceedings{huang2022layoutlmv3,
  title={Layoutlmv3: Pre-training for document ai with unified text and image masking},
  author={Huang, Yupan and Lv, Tengchao and Cui, Lei and Lu, Yutong and Wei, Furu},
  booktitle={Proceedings of the 30th ACM international conference on multimedia},
  pages={4083--4091},
  year={2022}
}

@inproceedings{santhanam2022colbertv2,
  title={Colbertv2: Effective and efficient retrieval via lightweight late interaction},
  author={Santhanam, Keshav and Khattab, Omar and Saad-Falcon, Jon and Potts, Christopher and Zaharia, Matei},
  booktitle={Proceedings of the 2022 Conference of the North American Chapter of the Association for Computational Linguistics: Human Language Technologies},
  pages={3715--3734},
  year={2022}
}

@article{babakhin2025llama,
  title={Llama-Embed-Nemotron-8B: A Universal Text Embedding Model for Multilingual and Cross-Lingual Tasks},
  author={Babakhin, Yauhen and Osmulski, Radek and Ak, Ronay and Moreira, Gabriel and Xu, Mengyao and Schifferer, Benedikt and Liu, Bo and Oldridge, Even},
  journal={arXiv preprint arXiv:2511.07025},
  year={2025}
}

@article{yang2025qwen3,
  title={Qwen3 technical report},
  author={Yang, An and Li, Anfeng and Yang, Baosong and Zhang, Beichen and Hui, Binyuan and Zheng, Bo and Yu, Bowen and Gao, Chang and Huang, Chengen and Lv, Chenxu and others},
  journal={arXiv preprint arXiv:2505.09388},
  year={2025}
}

@article{dubey2024llama,
  title={The llama 3 herd of models},
  author={Dubey, Abhimanyu and Jauhri, Abhinav and Pandey, Abhinav and Kadian, Abhishek and Al-Dahle, Ahmad and Letman, Aiesha and Mathur, Akhil and Schelten, Alan and Yang, Amy and Fan, Angela and others},
  journal={arXiv e-prints},
  pages={arXiv--2407},
  year={2024}
}

@article{lee2024nv,
  title={NV-Embed: Improved Techniques for Training LLMs as Generalist Embedding Models},
  author={Lee, Chankyu and Roy, Rajarshi and Xu, Mengyao and Raiman, Jonathan and Shoeybi, Mohammad and Catanzaro, Bryan and Ping, Wei},
  journal={arXiv preprint arXiv:2405.17428},
  year={2024}
}

@article{moreira2024enhancing,
  title={Enhancing Q\&A Text Retrieval with Ranking Models: Benchmarking, fine-tuning and deploying Rerankers for RAG},
  author={Moreira, Gabriel de Souza P and Ak, Ronay and Schifferer, Benedikt and Xu, Mengyao and Osmulski, Radek and Oldridge, Even},
  journal={arXiv preprint arXiv:2409.07691},
  year={2024}
}

@inproceedings{diaonemotron,
  title={Nemotron-CLIMB: Clustering-based Iterative Data Mixture Bootstrapping for Language Model Pre-training},
  author={Diao, Shizhe and Yang, Yu and Fu, Yonggan and Dong, Xin and Su, Dan and Kliegl, Markus and Chen, Zijia and Belcak, Peter and Suhara, Yoshi and Yin, Hongxu and others},
  booktitle={The Thirty-ninth Annual Conference on Neural Information Processing Systems Datasets and Benchmarks Track},
  year={2025}
}

@misc{gemini-embed,
      title={Gemini Embedding: Generalizable Embeddings from Gemini}, 
      author={Jinhyuk Lee and Feiyang Chen and Sahil Dua and Daniel Cer and Madhuri Shanbhogue and Iftekhar Naim and Gustavo Hernández Ábrego and Zhe Li and Kaifeng Chen and Henrique Schechter Vera and Xiaoqi Ren and Shanfeng Zhang and Daniel Salz and Michael Boratko and Jay Han and Blair Chen and Shuo Huang and Vikram Rao and Paul Suganthan and Feng Han and Andreas Doumanoglou and Nithi Gupta and Fedor Moiseev and Cathy Yip and Aashi Jain and Simon Baumgartner and Shahrokh Shahi and Frank Palma Gomez and Sandeep Mariserla and Min Choi and Parashar Shah and Sonam Goenka and Ke Chen and Ye Xia and Koert Chen and Sai Meher Karthik Duddu and Yichang Chen and Trevor Walker and Wenlei Zhou and Rakesh Ghiya and Zach Gleicher and Karan Gill and Zhe Dong and Mojtaba Seyedhosseini and Yunhsuan Sung and Raphael Hoffmann and Tom Duerig},
      year={2025},
      eprint={2503.07891},
      archivePrefix={arXiv},
      primaryClass={cs.CL},
      url={https://arxiv.org/abs/2503.07891}, 
}

@misc{embedding-gemma,
      title={EmbeddingGemma: Powerful and Lightweight Text Representations}, 
      author={Henrique Schechter Vera and Sahil Dua and Biao Zhang and Daniel Salz and Ryan Mullins and Sindhu Raghuram Panyam and Sara Smoot and Iftekhar Naim and Joe Zou and Feiyang Chen and Daniel Cer and Alice Lisak and Min Choi and Lucas Gonzalez and Omar Sanseviero and Glenn Cameron and Ian Ballantyne and Kat Black and Kaifeng Chen and Weiyi Wang and Zhe Li and Gus Martins and Jinhyuk Lee and Mark Sherwood and Juyeong Ji and Renjie Wu and Jingxiao Zheng and Jyotinder Singh and Abheesht Sharma and Divyashree Sreepathihalli and Aashi Jain and Adham Elarabawy and AJ Co and Andreas Doumanoglou and Babak Samari and Ben Hora and Brian Potetz and Dahun Kim and Enrique Alfonseca and Fedor Moiseev and Feng Han and Frank Palma Gomez and Gustavo Hernández Ábrego and Hesen Zhang and Hui Hui and Jay Han and Karan Gill and Ke Chen and Koert Chen and Madhuri Shanbhogue and Michael Boratko and Paul Suganthan and Sai Meher Karthik Duddu and Sandeep Mariserla and Setareh Ariafar and Shanfeng Zhang and Shijie Zhang and Simon Baumgartner and Sonam Goenka and Steve Qiu and Tanmaya Dabral and Trevor Walker and Vikram Rao and Waleed Khawaja and Wenlei Zhou and Xiaoqi Ren and Ye Xia and Yichang Chen and Yi-Ting Chen and Zhe Dong and Zhongli Ding and Francesco Visin and Gaël Liu and Jiageng Zhang and Kathleen Kenealy and Michelle Casbon and Ravin Kumar and Thomas Mesnard and Zach Gleicher and Cormac Brick and Olivier Lacombe and Adam Roberts and Qin Yin and Yunhsuan Sung and Raphael Hoffmann and Tris Warkentin and Armand Joulin and Tom Duerig and Mojtaba Seyedhosseini},
      year={2025},
      eprint={2509.20354},
      archivePrefix={arXiv},
      primaryClass={cs.CL},
      url={https://arxiv.org/abs/2509.20354}, 
}

@misc{model-merge-2018,
      title={Averaging Weights Leads to Wider Optima and Better Generalization}, 
      author={Pavel Izmailov and Dmitrii Podoprikhin and Timur Garipov and Dmitry Vetrov and Andrew Gordon Wilson},
      year={2019},
      eprint={1803.05407},
      archivePrefix={arXiv},
      primaryClass={cs.LG},
      url={https://arxiv.org/abs/1803.05407}, 
}

@misc{model-merge-2022,
      title={Model soups: averaging weights of multiple fine-tuned models improves accuracy without increasing inference time}, 
      author={Mitchell Wortsman and Gabriel Ilharco and Samir Yitzhak Gadre and Rebecca Roelofs and Raphael Gontijo-Lopes and Ari S. Morcos and Hongseok Namkoong and Ali Farhadi and Yair Carmon and Simon Kornblith and Ludwig Schmidt},
      year={2022},
      eprint={2203.05482},
      archivePrefix={arXiv},
      primaryClass={cs.LG},
      url={https://arxiv.org/abs/2203.05482}, 
}

@article{tibshirani2001estimating,
  title={Estimating the number of clusters in a data set via the gap statistic},
  author={Tibshirani, Robert and Walther, Guenther and Hastie, Trevor},
  journal={Journal of the royal statistical society: series b (statistical methodology)},
  volume={63},
  number={2},
  pages={411--423},
  year={2001},
  publisher={Wiley Online Library}
}

@inproceedings{ma2024unifying,
  title={Unifying multimodal retrieval via document screenshot embedding},
  author={Ma, Xueguang and Lin, Sheng-Chieh and Li, Minghan and Chen, Wenhu and Lin, Jimmy},
  booktitle={Proceedings of the 2024 Conference on Empirical Methods in Natural Language Processing},
  pages={6492--6505},
  year={2024}
}

@article{yu2024visrag,
  title={Visrag: Vision-based retrieval-augmented generation on multi-modality documents},
  author={Yu, Shi and Tang, Chaoyue and Xu, Bokai and Cui, Junbo and Ran, Junhao and Yan, Yukun and Liu, Zhenghao and Wang, Shuo and Han, Xu and Liu, Zhiyuan and others},
  journal={arXiv preprint arXiv:2410.10594},
  year={2024}
}

@article{deshmukh2025nvidia,
  title={Nvidia nemotron nano v2 vl},
  author={Deshmukh, Amala Sanjay and Chumachenko, Kateryna and Rintamaki, Tuomas and Le, Matthieu and Poon, Tyler and Taheri, Danial Mohseni and Karmanov, Ilia and Liu, Guilin and Seppanen, Jarno and Chen, Guo and others},
  journal={arXiv preprint arXiv:2511.03929},
  year={2025}
}

@String{Computer = "{IEEE} Computer" }

%%
%% If your work has an appendix, this is the place to put it.
\appendix

%\section{Appendix 1}

\end{document}